%% file: vu-corr.tex
\documentclass{article}
\usepackage[ruled,vlined,longend]{algorithm2e}
\input{jl_definitions}

\newcommand{\tup}[1]{{\ensuremath{\scriptstyle <}}#1{\ensuremath{\scriptstyle >}}}
\newcommand{\tago}[1]{{\ensuremath{\scriptstyle <}}#1{\ensuremath{\scriptstyle >}}}
\newcommand{\tagc}[1]{{\ensuremath{\scriptstyle <\hspace*{-.2em}/}}#1{\ensuremath{\scriptstyle >}}}

\newcommand{\co}{\hspace*{-.5ex}:\hspace*{-.5ex}}

\newcommand{\mm}{\hspace*{2em}}

\newcommand{\frake}{\ensuremath{\mathfrak{e}}}
\newcommand{\calL}{\ensuremath{{\cal L}}}
\newcommand{\barT}{\ensuremath{\bar{T}}}

\title{Update XML Views}

\author{Jixue Liu$^1$ \hspace{1cm} Chengfei Liu$^2$ \hspace{1cm}
   Theo Haerder$^3$ \hspace{1cm} Jeffery Xu Yu$^4$
   \\   $^1$ School of Computer and Information Science,
    University of South Australia \\
    \hspace*{1cm}email: \{jixue.liu\}@unisa.edu.au\\
    $^2$ Faculty of ICT, Swinburne University of Technology\\
    \hspace*{1cm}email: cliu@swin.edu.au \\
    $^3$ Dept of Computer Sciences, Technical University of
    Kaiserslautern \\
    \hspace*{1cm}email: haerder@informatik.uni-kl.de \\
    $^4$ Dept of Systems Eng. and Eng. Management, Chinese University of
    HK\\
    \hspace*{1cm}email: yu@se.cuhk.edu.hk \\
     \\
    Completed 2011 June
    }

\begin{document}
\date{}
\maketitle

\begin{abstract}
View update is the problem of translating an update to a view to some updates to the
source data of the view. In this paper, we show the factors determining XML view update
translation, propose a translation procedure, and propose translated updates to the
source document for different types of views. We further show that the translated updates
are precise. The proposed solution makes it possible for users who do not have access
privileges to the source data to update the source data via a view.
\end{abstract}



{\em keywords:} XML data, view update, update translation, virtual views

\section{Introduction}\label{sec:intro}

A (virtual) view is defined with a query over some {\em source data} of a database. The
query is called the {\em view definition} which determines what data appears in the view.
The data of the view, called a \emph{view instance}, is often not stored in the database
but is derived from the source data on the fly using the view definition every time when
the view is selected.

In database applications, many users do not have privileges to access all the data of a
database. They are often given a view of the database so that they can retrieve only the
data in the view. When these users need to update the data of the database, they put
their updates against the view, not against the source data, and expect that the view
instance is changed when it is accessed next time. This type of updates is called a
{\emph view update}. {\it Because of its important use}, view update has a long research
history
\cite{banc81tods-vu-seman,masu84vldb-vu-trans,tom88icdt-vu-anno,Tom94ddlp-vu-qc,falq00TR-htxt-vu,brag04vldb-xr-vu,wang06dke-xr-vu}.
The work in \cite{cong07DNIS-xmlview-review} discusses detailed semantics of view updates
in many scenarios.

Unfortunately, view updates cannot be directly applied to the view instance as it is not
stored physically  and is derived on the fly when required (virtual view). Even in the
cases where the view instance is stored (materialized view), which is not the main focus
of this paper, applying updates to the instance may cause inconsistencies between the
source data and the instance. To apply a view update to a virtual view, a translation
process is required to translate the view update to some {\it source updates}. When the
source data is changed, the data in the view will be changed next time when the view is
selected. To the user of the view, it seems that the view update has been successfully
applied to the view instance.

Let $V$ be a view definition, $V^i$ the view instance, $S^i$ the source data of the view,
$V(S^i)$ the evaluation of $V$ against $S^i$. Then $V^i=V(S^i)$. Assume that the user
wants to apply a view update $\delta V$ to $V^i$ as $\delta V(V^i)$. View update
translation is to find a process that takes $V$ and $\delta V$ as input and produces a
source update $\delta S$ to $S^i$ such that next time when the user accesses the view,
the view instance appears changed and is as expected by the user. That is, for any $S^i$
and $V^i=V(S^i)$,

\begin{equation}\label{eq:vu-requ} V(\delta S(S^i))=\delta V(V^i) \end{equation}

Two typical anomalies, view side-effect and source document over-update, are easily
introduced by the translation process although they are update policy dependent
\cite{masu84vldb-vu-trans}. View side-effect \cite{wang06dke-xr-vu} is the case where the
translated source update causes more-than-necessary change to the source data which leads
to more-than-expected change to the view instance. View side-effect makes Equation
(\ref{eq:vu-requ}) violated.

Over-updates may also happen to a source document. An over-update to a source document
causes the source data irrelevant to the view to be changed, but keeps the equation
satisfied. A source document over-update is incorrect as it changes information that the
user did not expect to change.

A \emph{precise} translation of a view update should produce source updates that (1)
result in necessary (as the user expects) change to the view instance, (2) do not cause
view side-effect, and (3) do not cause over-updates to the source documents.

In relational databases, extensive work has been done on view update and the problem has
been well understood \cite{banc81tods-vu-seman,masu84vldb-vu-trans,tom88icdt-vu-anno}. In
cases of updating XML views over relational databases, updates to XML views need to be
translated to updates to the base relational tables. The works in
\cite{brag04vldb-xr-vu,wang06dke-xr-vu} propose two different approaches to the problem.
The work in \cite{brag04vldb-xr-vu} translates an XML view to some relational views and
an update to the XML view to updates to the relational views. It then uses the relational
approach to derive updates to the base tables. The work in \cite{wang06dke-xr-vu} derives
a schema for the XML view and annotates the schema based on keys of relational tables and
multiplicities. An algorithm is proposed to use the annotation to determine if a
translation is possible and how the translation works. Both works assume keys, foreign
keys and the join operator based on these two types of constraints. Another work,
technical report \cite{falq00TR-htxt-vu}, proposes brief work on updating hypertext views
defined on relational databases. To the best of our knowledge, the only work relating to
XML view update is \cite{liu07pepm-Xvu-biDire} which proposes a middle language and a
transformation system to derive view instance from source data, and to derive source data
from a {\bf materialized} view instance, and assumes XQuery as the view definition
language. We argue that with the view update problem, only view updates are available but
not the view instance (not materialized). Consequently view update techniques are still
necessary.

In this paper, we look into the view update problem in pure XML context. This means that
both source data and the view are in XML format. We assume that base XML documents have
no schema and no constraints information available.

The view update problem in the relational database is already difficult as not all view
updates are translatable. For example, if a view $V$ is defined by a Cartesian product of
two tables $R$ and $S$, an update inserting a new tuple to the view instance is not
translatable because there is no unique way to determine the change(s) to $R$ and $S$.
The view update problem in XML becomes much harder. The main reason is that the source
data and view instances are modeled in trees and trees can nest in arbitrary levels. This
fundamental difference makes the methods of translating view updates in the relational
database not applicable to translating XML view updates. For example, the selection and
the projection in the relational database do not have proper counterparts in XML. The
view update problem in XML has many distinct cases that do not exist in the view update
problem in the relational database (see Sections \ref{sec:translation1} and
\ref{sec:translation3} for details). To the best of our knowledge, our work is the first
proposing a solution to the view update problem in XML.

We notice that the view update problem is different from the view maintenance problem.
The former aims to translate a view update to a virtual view to a source update while the
latter aims to translate a source update to a view update to a materialized view. The
methods for one do not work for the other.

We make the following contributions in this paper. Based on the view definition and the
update language presented later, we identify the factors determining the view update
problem. We propose a translation algorithm to translate view updates to source updates.
Furthermore, we propose translated updates to the source for different types of view
updates. The types of view updates range from the case where the update involves an
individual tree selected the source, the case where the update involves multiple trees
from the source, and the case where the update happens to the root of the view. For each
proposed update to the source, we prove that it is precise.

The paper is organized as follows. Section \ref{sec:model} shows the view definition
language, the update language, and the preciseness of view update translation. In Section
\ref{sec:translation1}, we propose an algorithm and show that the translation obtained by
the algorithm is a precise translation. In Section \ref{sec:translation2}, we identify a
`join' case where a translated update is precise. Section \ref{sec:translation3} shows a
translation when a main subtree of the view is deleted. Section \ref{sec:conclusion}
concludes the paper.

 \vspace{-1ex}
\section{Preliminaries} \label{sec:model}

 \vspace{-1ex}
In this section, we define basic notation, introduce the languages for view definitions
and updates, and define the XML view update problem.

 \begin{defi}[tree]
An XML document can be represented as an ordered tree. Each node of the tree has a unique
identifier $v_i$, an element name $ele$ also called a \emph{label}, and either a text
string $txt$ or a sequence of child trees $T_{j_1}, \ddd, T_{j_n}$. That is, a node is
either $(v_i:ele:txt)$ or $(v_i:ele:T_{j_1},\ddd, T_{j_n})$. When the context is clear,
some or all of the node identifiers of a tree may not present explicitly. A tree without
all node identifiers is called a \emph{value tree}. Two trees $T_1$ and $T_2$ are (value)
\emph{equal}, denoted by $T_1 = T_2$, if they have identical value trees. If a tree $T_1$
is a subtree in $T_2$, $T_1$ is said \emph{in} $T_2$ and denoted by $T_1\in T_2$. \done
 \end{defi}

For example, the document {\tt <root><A><B>1</B></A><A><B>2</B></A></root>} is
represented by $T=(v_r\co root \co \ (v_0 \co A \co (v_1 \co B \co 1)), (v_2 \co A \co
(v_3 \co B \co 2)))$. The value tree of $T$ is $( root \co \ ( A \co ( B \co 1)),( A \co
( B \co 2)))$.

 \vspace{-1ex}
\begin{defi}\label{def:path}
A \emph{path} $p$ is a sequence of element names $e_1/e_2/\ddd/e_n$ where all names are
distinct. The function $L(p)$ returns the last element name $e_n$.

Given a path $p$ and a sequence of nodes $v_1,\ddd,v_n$ in a tree, if for every node
$v_i\in[v_2,\ddd,v_n]$, $v_i$ is labeled by $e_i$ and is a child of $v_{i-1}$, then
$v_1/\ddd/v_{n}$ is a \emph{doc path} conforming to $p$ and the tree rooted at $v_n$ is
denoted  by $T^p_{v_n}$. \done
\end{defi}

\subsection{View definition language}

We assume that a view is defined in a dialect of the $for$-$where$-$return$ clauses of
XQuery \cite{xquery1-boag07}.

\begin{defi}[$V$] A view is defined by
 \vspace*{-1ex}
\begin{Verbatim}[commandchars=\\\{\},codes={\catcode`$=3\catcode`^=7\catcode`_=8}]
 \tago{$v$}\{ for $x_1$ in $p_1$,   $\ddd$,  $x_n$ in $p_n$
       where  $cdn(x_1,\ddd,x_n)$
       return $rtn(x_1,\ddd,x_n)$  \}\tago{$/v$}
\end{Verbatim}
 where $p_1,\ddd,p_n$ are paths (Definition \ref{def:path}) proceeded by $doc()$ or $x_i$;\\
 \mm $cdn(x_1,\ddd,x_n)$ ::= $x_i/{\cal E}_i=x_j/{\cal E}_j$ and $\ddd$ and  $x_k/{\cal E}_k=strVal$ and $\ddd$; \\
 \mm $rtn(x_1,\ddd,x_n)$ ::= \tago{$\mathfrak{e}$} \{$x_u/\gamma_u$\} $\ddd$ \{$x_v/\gamma_v$\} \tagc{$\mathfrak{e}$}; \\
 \mm $\gamma,{\cal E}$ are paths,
   and the last elements of all $x_u/\gamma_u,\ddd,x_v/\gamma_v$ are distinct.  \done
 \end{defi}

  \vspace{1ex}
We note that the paths in the $return$ clause are denoted by $x_i/\gamma$s because these
expressions are specially important in view update translation. We purposely leave out
the \$ sign proceeding a variable in the XQuery language.

\begin{defi}[context-based production] \label{def:cb-product}
By the formal semantics of XQuery \cite{Frank01sigmodRec-XQseman}, the semantics of
the language is
 \vspace*{-1ex}
\begin{Verbatim}[commandchars=\\\{\},codes={\catcode`$=3\catcode`^=7\catcode`_=8}]
   for $x_1$ in $p_1$ return
      for $x_2$ in $p_2$ return
         ...
            for $x_n$ in $p_n$ return
               if cdn($x_1,...,x_n$)=true
                  return rtn($x_1,...,x_n$)
\end{Verbatim}
The for-statement produces tuples $\tup{x_1,...,x_n}$, denoted by $fortup(V)$,  where the
variable $x_i$ represents a binding out of the sub trees located by $p_i$ within the
context defined by $x_{1},\ddd,x_{i-1}$. This process is called \emph{context-based
production}. \done
\end{defi}

For each tuple satisfying the condition $cdn(x_1,...,x_n)$, the function
$rtn(x_1,\ddd,x_n)$ produces a tree, called an $\mathfrak{e}$-tree, under the root node
of the view. That is, $V$ maps a tuple to an $\frake$-tree. The children of the
$\mathfrak{e}$-tree are the \emph{$\gamma$-trees} selected by all the expressions
$x_i/\gamma_i$s (for all $i$) from the tuple. A tuple is mapped to one and only one
$\frake$-tree and an $\frake$-tree is for one and only one tuple. A $\gamma$-tree of a
tuple is uniquely mapped to a child of the $\frake$-tree of the tuple and  a child of an
$\frake$-tree is for one and only one $\gamma$-tree of its tuple.

The path of a node $s$ in the view has the following format:

   \begin{equation}\label{eq:vpath} v/\mathfrak{e}/\calL_i/\theta_i \end{equation}
   \vspace{-1em}
   \begin{equation}\label{eq:mapping}\calL_i=L(x_i/\gamma_i)\end{equation}

where $x_i/\gamma_i$ is an expression in $rtn(x_1,...,x_n)$, $L(x_i/\gamma_i)$ returns
the last element name $\calL_i$ of the path $x_i/\gamma_i$, and $\theta_i$ is a path
following $\calL_i$ in the view. When $\calL_i/\theta_i$ is not empty, the path in the
source document corresponding to $v/\mathfrak{e}/\calL_i/\theta_i$ is

   \[ x_i/\gamma_i/\theta_i \]

The view definition has some properties important to view update translation. Firstly
because of context-based production, a binding of variable $x_i$ may be copied into
$x^{(1)}_i,\ddd, x^{(m)}_i$ to appear in multiple tuples:
 \[\tup{\ddd,x^{(1)}_i,\ddd,x_{j[1]},\ddd}\]
   \vspace{-2em}
 \[\ddd \hspace*{10em}\]
   \vspace{-2em}
 \[\tup{\ddd,x^{(m)}_i,\ddd,x_{j[m_j]},\ddd}\]
where $x_{j[1]}, \ddd, x_{j[m_j]}$ are different bindings of $x_j$. Each tuple satisfying
the condition $cdn(x_1,\ddd,x_n)$ is used to build an $\mathfrak{e}$-tree. As a result of
$x_i$ being copied, the subtrees of $x_i$ will be copied accordingly to appear in
multiple $\mathfrak{e}$-trees in the view.

Secondly, a tree may have zero or many sub trees located by a given path $p$. That is,
given a tree bound to $x_i$, the path expression $x_i/p$ may locate zero or many sub
trees $T^{x_i/p}_1,\ddd, T^{x_i/p}_{n_p}$  in $x_i$. This is true in the source documents
and in the view.

Thirdly, two path expressions $x_i/\gamma_i$ and $x_j/\gamma_j$ generally may have the
same last element name, i.e., $L(x_i/\gamma_i)=L(x_j/\gamma_j)$. For example, if $x_i$
represents an employee while $x_j$ represents a department, then $x_i/name$ and
$x_j/name$ will present two types of names in the same $\mathfrak{e}$-tree. This make the
semantics of the view data not clear. This is the reason that we assume that all
$L(x_i/\gamma_i)$s are distinct.

\begin{exam}\label{ex:view-def}
Consider the view definition below and the source document shown in Figure
\ref{f:running}(a). The view instance is shown in Figure \ref{f:running}(b).

 \vspace{-1ex}
\label{pg:viewQ}
\begin{Verbatim}[commandchars=\\\{\},codes={\catcode`$=3\catcode`^=7\catcode`_=8}]
 <v>\{for x in doc("r")/r/A,  y in x/C,   z in x/H
      where y/D=z and z="1"
      return <$\mathfrak{e}$>\{x/B\}\{x/C\}\{y/F/G\}\{z\}</$\mathfrak{e}$>
  \}</v>
\end{Verbatim}

\begin{figure*}[h] \center
      \includegraphics[scale=0.8]{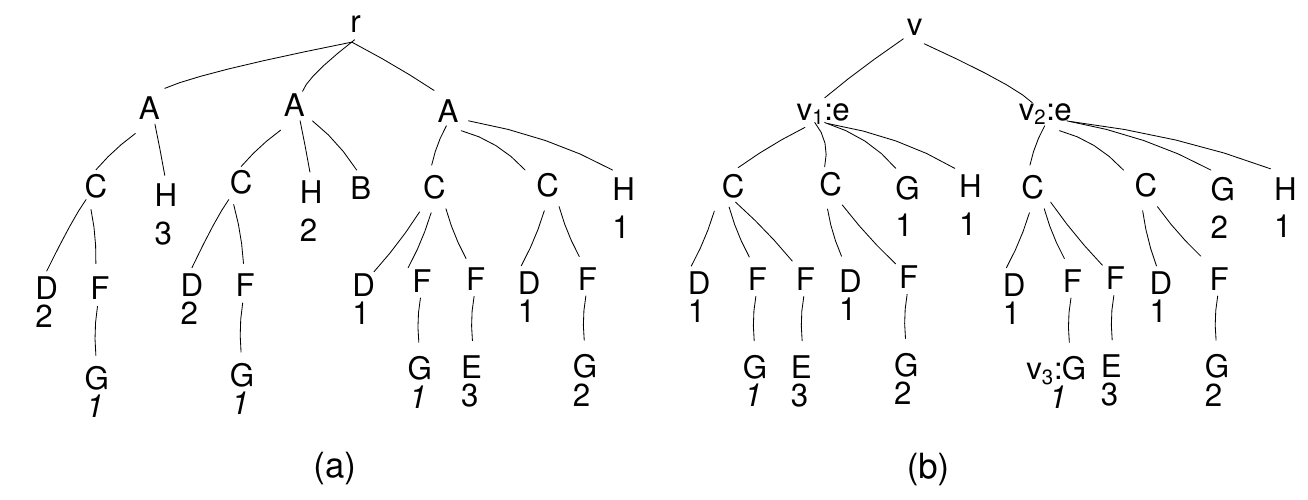}
     \caption{Source document $r$ and view $v$}
   \label{f:running}
\end{figure*}

 \vspace{-1ex}
 From the view definition, $\gamma_1=B$, $\gamma_2=C$, $\gamma_3=F/G$, and
$\gamma_4=\phi$. $L(x/\gamma_1)=\calL_1=B$, $L(x/\gamma_2)=\calL_2=C$,
$L(y/\gamma_3)=\calL_3=G$, and $L(z/\gamma_4)=\calL_4=H$.

Formula (\ref{eq:vpath}) is exemplified as the following. The node $v_3$ in the view has
the path $v/e/C/F/G$ where $C$ is $\calL_2=L(x/\gamma_2)$ and $F/G$ is $\theta$. The node
$v_1$ is an $\mathfrak{e}$ node and its path is $v/e$ and $\calL_i/\theta_i$ is $\phi$.

The example shows the following.
\begin{itm}
 \item
The expression $x/B$ (=$x/\gamma_1$) of the $return$ clause has no tree in the
$\mathfrak{e}$-trees.
 \item
The path expression $x/C$ (=$x/\gamma_2$) has multiple trees in an $\mathfrak{e}$-tree.
 \item
The trees of $x/C$ are duplicated in the view and so are their sub trees.
 \item
Each of some $x/C$ trees has more than one $x/C/F$ (=$x/\gamma_2/\theta$) sub trees.
\end{itm}
\end{exam}

 \vspace*{-2em}
\subsection{The update language}

The update language we use follows the proposal \cite{tata01sigmod-ux} extended from
XQuery.

\begin{defi}[$\delta V$] A view update statement has the format of
 \vspace*{-.5em}
\begin{Verbatim}[commandchars=\\\{\},codes={\catcode`$=3\catcode`^=7\catcode`_=8}]
 for $\bar{x}_1$ in $\bar{p}_1$,   $\ddd$,   $\bar{x}_u$ in $\bar{p}_u$
 where $\bar{x}_c/\bar{p}_{c}=strValu$
 update $\bar{x}_t/\bar{p}_t$ ( delete $T$ |  insert $T$ )
\end{Verbatim}
where $\bar{x}_c,\bar{x}_t \in [\bar{x}_1,\ddd,\bar{x}_u]$, $\bar{p}_1,\ddd, \bar{p}_u$
are paths (Definition \ref{def:path}) proceeded by $v$ or $\bar{x}_i$;
$\bar{p}_c,\bar{p}_t$ are paths; all element names in the paths are elements names in the
view. $\bar{x}_c/\bar{p}_{c}$ and $\bar{x}_t/\bar{p}_t$ are called the \emph{(update)
condition path} and \emph{(update) target path} respectively. \done
\end{defi}

The next process builds the mapping represented by Formula (\ref{eq:mapping}).
 \vspace*{-1ex}
\begin{proc}[mapping] \label{def:mapping}  When the variables in $\bar{x}_c/\bar{p}_c$ and
$\bar{x}_t/\bar{p}_t$ are replaced by their paths in the $for$-clause until the first
element name becomes $v$, the full paths of $\bar{x}_c/\bar{p}_c$ and
$\bar{x}_t/\bar{p}_t$ will have the format of $v/\mathfrak{e}/{\cal L}_c/\theta_c$ and
$v/\mathfrak{e}/{\cal L}_t/\theta_t$ as shown in Formula (\ref{eq:vpath}). The element
names $\calL_c$ and $\calL_t$, if $\calL_c/\theta_c$ and ${\cal A}_t/\theta_t$ are not
empty, must be the last element names of two expressions $x_c/\gamma_c$ and
$x_t/\gamma_t$ in the $return$ clause of the view definition $V$. A search using
$\calL_c$ and $\calL_t$ in $V$ will identify the expressions. Consequently
$v/\mathfrak{e}/\calL_c/\theta_c$ and
  $v/\mathfrak{e}/\calL_t/\theta_t$ are mapped to $x_c/\gamma_c/\theta_c$ and
  $x_t/\gamma_t/\theta_t$ respectively. \done
\end{proc}

With this mapping, the update statement $\delta V$ can be represented by the following
abstract form:
 \vspace*{-1ex}
 \begin{equation}
 (\bar{p}_s; \ \ v/\mathfrak{e}/{\cal L}_c/\theta_c=strValu; \ \ v/\mathfrak{e}/{\cal L}_t/\theta_t;  \ \ del(T)|ins(T) )
 \end{equation}

 \vspace*{-1em}
where
 \begin{itm}
 \item
 $v/\mathfrak{e}/{\cal L}_c/\theta_c$ is the \emph{full} update condition path
(int the view) for $\bar{x}_c/\bar{p}_{c}$, $v/\mathfrak{e}/{\cal L}_t/\theta_t$ the
\emph{full} target path for $\bar{x}_t/\bar{p}_{t}$;
 \item
 $\bar{p}_s$ is the maximal common front part of $v/\mathfrak{e}/{\cal L}_c/\theta_c$
 and $v/\mathfrak{e}/{\cal L}_t/\theta_t$.
 \end{itm}

The semantics of an update statement is that under a context node identified by
$\bar{p}_s$, if a sub tree identified by  $v/\mathfrak{e}/{\cal L}_c/\theta_c$ satisfies
the update condition, all the sub trees identified by $v/\mathfrak{e}/{\cal
L}_t/\theta_t$ will be applied the update action (del(T) or ins(T)). The sub tree
$T^{v/\mathfrak{e}/{\cal L}_c/\theta_c}$ is called the \emph{condition tree} of
$T^{v/\mathfrak{e}/{\cal L}_t/\theta_t}$. A sub tree is updated only if it has a
condition tree and the condition tree satisfies the update condition. An update target
and its condition trees are always within a tuple when the view definition is evaluated
and are in an $\frake$-tree in the view after the evaluation.

We note that because of the context-based production in the update language, the same
update action may be applied to a target node for multiple times. For example, if $x$ is
binding and the context-based production produces two tuple for it $<x^{(1)},\ddd>$ and
$<x^{(2)},\ddd>$. If the update condition and target are all in $x$, $x$ will be updated
twice with the same action. We assume that only the effect of the first application is
taken and the effect of all other applications are ignored.

 \vspace{1ex}
Based on the structure of the target path $tp=v/\mathfrak{e}/{\cal L}_t/\theta_t$,
updates may happen to different types of nodes in the view.
 \begin{itm}
  \item
  When ${\cal L}_t/\theta_t \not = \phi$, the update happens to the nodes within a $\gamma$-tree.
  \item
 When $tp=v/\mathfrak{e}$, the update will add or delete a $\gamma$-tree.
 \item
 When $tp=v$ (in this case, $\bar{p}_s=v$), the update will add or delete an $\frake$-tree.
\end{itm}

We will present the first case in Sections \ref{sec:translation1} and
\ref{sec:translation2} and present the last two cases in Section \ref{sec:translation3}.

 \vspace*{-1em}
\subsection{The view update problem}

\begin{defi}[Precise Translation]\label{def:precise}
Let $V$ be a view definition and $S$ be the source of $V$. Let $\delta V$ be an update
statement to $V$. Let $\delta S$ be the update statement to $S$ translated from $\delta
V$. $\delta S$ is a precise translation of $\delta V$ if, for any instance $S^i$ of $S$
and $V^i=V(S^i)$,
  \begin{list1}
  \item  $\delta S$ is correct. That is,
$V(\delta S(S^i))==\delta V(V^i)$ is true; and
  \item  $\delta S$ is minimal. That is, there does not exist another translation
  $\delta S'$ such that ($\delta S'$ is correct, i.e.,
  $V(\delta S'(S^i))=V(\delta S(S^i))=\delta V(V^i)$ and there exists a tree $T$ in $S^i$ and $T$ is
  updated by $\delta S$ but not $\delta S'$). \done
\end{list1}
\end{defi}
We note that Condition (1) also means that the update $\delta S$ will not cause
view-side-effect. Otherwise, $V(\delta S(S^i))$ would contain more, less, or different
updated trees than those in $\delta V(V^i)$.


\begin{defi}[the view  update problem] Given a view $V$ and a view update $\delta V$,
the problem of view update is to (1) develop a translation process $P$, and show that the
source update $\delta S$ obtained from $P$ is precise, or (2) prove that a precise
translation of $\delta V$ does not exist. \done
\end{defi}

\section{Update Translation when ${\cal L}_t/\theta_t \not=\phi$ and $x_c=x_t$} \label{sec:translation1}
In this section, we investigate update translation when the update is to change a
$\gamma$-tree of the view and the mappings of the update condition path and the target
path refer to the same variable. We present Algorithm \ref{alg:translate} for view update
translation in this case. The algorithm is self-explainable.

\begin{algorithm}  \label{alg:translate}
 \caption{A translation algorithm }
   \LinesNumbered 
 \KwIn{view definition $V$, view update $\delta V$}
 \KwOut{translated source update $\delta S$}
 \Begin{
   make a copy of $V$ and reference the copy by $\delta S$ \;
   remove $rtn()$ from $\delta S$ \;
   from the view update $\delta V$, following Procedure \ref{def:mapping},
   find mappings $x_c/\gamma_c/\gamma_c$ and $x_t/\gamma_t/\gamma_t$ for the
   condition path $\bar{x}_c/\bar{p}_c$ and the target path $\bar{x}_t/\bar{p}_t$ \;
   make a copy of $\delta V$ and reference the copy by $\delta V_c$ \;
   in $\delta V_c$, replace $\bar{x}_c/\bar{p}_c$ and $\bar{x}_t/\bar{p}_t$ by
        $x_c/\gamma_c/\gamma_c$ and $x_t/\gamma_t/\gamma_t$ respectively \;
   append the condition in the $where$ clause of $\delta V_c$ to the end of the $where$
     clause in $\delta S$ using logic $and$ \;
   append the $update$ clause of $\delta V_c$ after the $where$ clause of
     $\delta S$
 }
\end{algorithm}

By the algorithm, the following source update is derived.

 \vspace*{-.5ex}
 \begin{equation}\label{eq:deltaS}\end{equation}
 \vspace*{-4em}
\begin{Verbatim}[commandchars=\\\{\},codes={\catcode`$=3\catcode`^=7\catcode`_=8}]
 $\delta$$S$:  for $x_1$ in $p_1$, \ \  $\ddd$, \ \ $x_n$ in $p_n$
       where $cdn(x_1,\ddd,x_n)$ and $x_c/\gamma_c/\theta_c=strValu$
       update $x_t/\gamma_t/\theta_t$  (insert $T$ | delete $T$)
 \end{Verbatim}

We now develop the preciseness of the translation. We recall notation that $fortup(V)$
means the tuples of the context-based production (Definition \ref{def:cb-product}) of
$V$. $x_c^{(1)}$ and $x_c^{(2)}$ are two copies of a binding of $x_c$, and $x_c$,
$x_{c[1]}$ and $x_{c[2]}$ are three separate bindings of $x_c$.

\begin{lemm}\label{lm:all-chged-in-tuple}
Given a tuple $t=<x_t,x_c,\ddd> \in fortup(V)$ and its $\frake$-tree $e$, (1) if $T$ is a
tree for the path $x_t/\gamma_t/\theta_t$ in $t$ and $T$ is updated by $\delta S$, then
all the trees identified by $x_t/\gamma_t/\theta_t$ in $t$ are updated by $\delta S$, and
all the trees identified by $\calL_t/\theta_t$ in $e$ are updated by $\delta V$. (2) if
$T$ is a tree for the path $\calL_t/\theta_t$ in $e$ and $T$ is updated by $\delta V$,
then all the trees identified by $x_t/\gamma_t/\theta_t$ in $t$ are updated by $\delta
S$, and all the trees identified by $\calL_t/\theta_t$ in $e$ are updated by $\delta V$.
\end{lemm}

The lemma is correct because of the one-to-one correspondences between a tuple and an
$\frake$-tree and between $t$'s $\gamma$-trees and $e$'s children, and because all the
trees identified by $x_t/\gamma_t/\theta_t$ in $t$ share the same condition tree(ies)
identified by $x_c/\gamma_c/\theta_c$ in $x_c$ of $t$, and all the trees identified by
$\calL_t/\theta_t$ in $e$ share the same condition tree(ies) identified by
$\calL_c/\theta_c$ in $e$.

\begin{lemm}\label{lm:noChange2cdn}
Given a tuple $t=<x_t,x_c,\ddd> \in fortup(V)$, let a subtree $T^{x_t/\gamma_t/\theta_t}$
of $x_t$ be updated by $\delta S$ and become $t'=<x'_t,x_c,\ddd>$. If
$x_t/\gamma_t/\theta_t$ is not a prefix of any of the path in the $where$ clause of
$\delta S$, if $t$ satisfies $cdn()$ of $V$, $t'$ also satisfies $cdn()$ of $V$.
\end{lemm}

The lemma is correct because the subtrees in the tuple used to test $cdn()$ are not
changed by $\delta S$ when the condition of the lemma is met.

\begin{lemm}\label{lm:updCondi-VS-all-true}
Given a tuple $t=<x_t,x_c,\ddd> \in fortup(V)$ and its $\frake$-tree $e$, if the
$T^{x_c/\gamma_c/\theta_c}$ in $t$ satisfies $x_c/\gamma_c/\theta_c = strValu$,
$T^{\calL_c/\theta_c}$ in $e$ satisfies $\calL_c/\theta_c = strValu$ and vice versa.
\end{lemm}

The correctness of the lemma is guaranteed by the one-to-one correspondence between $t$'s
$\gamma$-trees and $e$'s children.

\begin{lemm}\label{lm:VS-chged2same}
Given a tuple $t=<x_t,x_c,\ddd> \in fortup(V)$ and its $\frake$-tree $e$, let $T$ be a
tree identified by $x_t/\gamma_t/\theta_t$ in $t$ and $T'$ be the corresponding tree
identified by $\calL_t/\theta_t$ in $e$. Obviously $T=T'$. As $\delta S$ and $\delta V$
have the same update action, if $x_c$ satisfies the update condition, $\delta S(T)=\delta
V(T')$.
\end{lemm}

\begin{theo}\label{th:typic-precise}
Update $\delta S$ is a precise translation of the view update $\delta V$ if (i) ${\cal
L}_t/\theta_t \not=\phi$ and $x_c=x_t$, and (ii) $x_t/\gamma_t/\theta_t$ does not proceed
any path in the $where$ clause of $\delta S$.
\end{theo}

\newcommand{\Ttr}{\ensuremath{T^{x_1/\gamma_t}}}  
\newcommand{\Tcr}{\ensuremath{T^{x_1/\gamma_c}}}  
\newcommand{\Tts}{\ensuremath{T^{x_1/\gamma_t/\theta_t}}}  
\newcommand{\Ttv}{\ensuremath{T^{\calL_t/\theta_t}}}  
\newcommand{\Tcs}{\ensuremath{T^{x_1/\gamma_c/\theta_c}}}  
\newcommand{\Tcv}{\ensuremath{T^{\calL_c/\theta_c}}}  

\begin{figure*}[h] \center
      \includegraphics[scale=0.8]{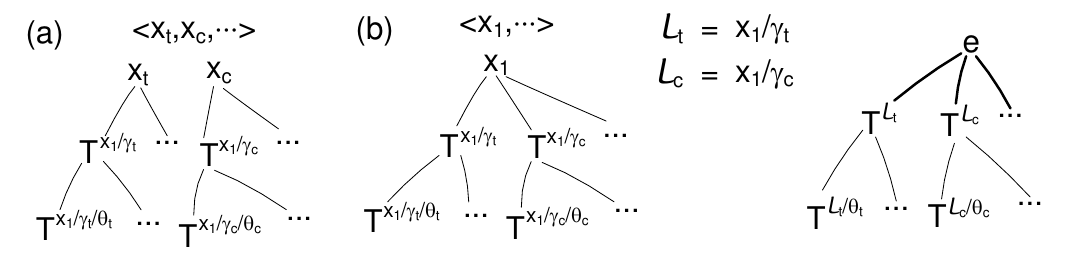}
     \caption{Each of tuples is mapped to an $\frake$-tree}
   \label{f:tup-e}
\end{figure*}

\noindent \bfit{Proof.} We follow Definition \ref{def:precise}. Without losing
generality, we assume that $x_t=x_c=x_1$. Figure \ref{f:tup-e} illustrates the
relationship between a variable binding $x_1$ in the tuple $<x_1,\ddd>$ and the
$\frake$-tree built from the tuple. The $\gamma$-trees $T^{x_1/\gamma_t}$ and
$T^{x_1/\gamma_c}$ in $x_1$ become the children of $e$ in the view. $\Tts$ and
 \Tcs are an update target tree and a condition tree respectively. $\Tts$'s children will
 be deleted or a new child will be inserted.

(1) Correctness: $V(\delta S(S^i))=\delta V(V(S^i))$

Consider two tuples $t_1=<x_1^{(1)},\ddd>$ and  $t_2=<x_1^{(2)},\ddd>$ in the evaluation
of $\delta S$ where $x_1^{(1)}$ and $ x_1^{(2)}$ are copies of $x_1$. Obviously if
$x_1^{(1)}$ is updated, $x_1^{(2)}$ is updated too. That is, their source $x_1$ will be
updated twice although only the first is effective. As $\delta S$ and $V$ have the same
$for$ clause, $t_1$ and $t_2$ exist in $fortup(V)$. Assume $e_1$ and $e_2$ are mapped
from $t_1$ and $t_2$ respectively by $V$. Then, either both $e_1$ and $e_2$ are updated
or none is updated.

 $\supseteq$: Let $T^{\calL_t/\theta_t}$ be a tree in an $\frake$-tree $e$ of
$V(S^i)$ updated to $\barT^{\calL_t/\theta_t}$ by $\delta V$ ($e$ becomes $e'$ after the
update). We show that $\barT^{\calL_t/\theta_t}$ is in $e'$ of $V(\delta S(S^i))$. In
fact, that $T^{\calL_t/\theta_t}$ is in $V(S^i)$ means that there exists one and only one
tuple $t=\tup{x_1,\ddd}$ in $fortup(V)$ satisfying $cdn()$, that in the tuple,
$x_1/\gamma_t/\theta$ identifies the source tree $T^{x_1/\gamma_t/\theta_t}$ of
$T^{\calL_t/\theta_t}$. \ $T^{\calL_t/\theta_t}$ being updated by $\delta V$ means that
there exists a condition tree $T^{{\cal L}_c/\theta_c}$ in $e$ and the condition tree
satisfies $v/\mathfrak{e}/{\cal L}_c/\theta_c=strValu$.

On the other side, because $V$ and $\delta S$ have the same $for$ clause, $t$ is in
$fortup(\delta S)$. Because $T^{{\cal L}_c/\theta_c}$ makes $v/\mathfrak{e}/{\cal
L}_c/\theta_c=strValu$ true, so $T^{x_1/\gamma_c/\theta_c}$ makes
$x_1/\gamma_c/\theta_c=strVal$ true (Lemma \ref{lm:updCondi-VS-all-true}). This means
$T^{x_1/\gamma_t/\theta_t}$ is updated by $\delta S$ and becomes
$\hat{T}^{x_1/\gamma_t/\theta_t}$. Thus $t$ becomes $t'=<\bar{x}_1,\ddd>$. Because of
Lemma \ref{lm:VS-chged2same},
$\barT^{x_1/\gamma_t/\theta_t}$=$\hat{T}^{x_1/\gamma_t/\theta_t}$ . Because of (ii) of
the theorem and Lemma \ref{lm:noChange2cdn}, $t'$ satisfies $cdn()$ and generalizes $e'$
in the view. So $\barT^{\calL_t/\theta_t}$ is in $V(\delta S(S^i))$.

 $\subseteq$: Let $T^{\calL_t/\theta_t}_1$ and $T^{\calL_t/\theta_t}_2$
be two trees in $V(\delta S(S^i))$ and their source tree(s) are updated by $\delta S$. We
show that $T^{\calL_t/\theta_t}_1$ and $T^{\calL_t/\theta_t}_2$ are in $\delta
V(V(S^i))$. There are three cases: (a) $T^{\calL_t/\theta_t}_1$ and
$T^{\calL_t/\theta_t}_2$ share the same source tree $T^{x_1/\gamma_t/\theta_t}$ (they
must appear in different $\frake$-trees in the view), and (b) $T^{\calL_t/\theta_t}_1$
and $T^{\calL_t/\theta_t}_2$ have different source trees $T^{x_1/\gamma_t/\theta_t}_1$
and $T^{x_1/\gamma_t/\theta_t}_2$. Case (b) has two sub cases: (b.1)
$T^{\calL_t/\theta_t}_1$ and $T^{\calL_t/\theta_t}_2$ appear in the same $\frake$-tree in
the view, and (b.2) $T^{\calL_t/\theta_t}_1$ and $T^{\calL_t/\theta_t}_2$ appear in
different $\frake$-trees.

Case (a): That $T^{x_1/\gamma_t/\theta_t}$ is updated by $\delta S$ means that there
exist two tuples $\tup{x_1^{(1)},\ddd}$ and $\tup{x_1^{(2)},\ddd}$ in $fortup(\delta S)$
such that  $x_1^{(1)}=x_1^{(2)}$, both tuples satisfy $cdn()$, and there exists condition
tree $T^{x_1/\gamma_c/\theta_c}$ in each tuple satisfying
$x_c/\gamma_c/\theta_c=strValu$, $T^{x_1/\gamma_t/\theta_t}$ is updated to
$\barT^{x_1/\gamma_t/\theta_t}$ by $\delta S$ (two update attempts with the same action
for the two tuples, only the effect of the first attempt is taken). After the update, the
tuples become $t'_1=\tup{\bar{x}_1^{(1)},\ddd}$ and $t'_2=\tup{\bar{x}_1^{(2)},\ddd}$. By
Lemma \ref{lm:noChange2cdn}, $t'_1$ and $t'_2$ satisfy $cdn$ of $V$ and produce $e_1,e_2
\in V(\delta S(S^i))$ and $\barT^{\calL_t/\theta_t}_1\in e_1$ and
$\barT^{\calL_t/\theta_t}_2\in e_2$.

On the other side, when $V$ is evaluated against $S^i$, $x_1$ is copied to two tuples
$t_1=\tup{x_1^{(1)},\ddd}$ and $t_2=\tup{x_1^{(2)},\ddd}$ in $fortup(V)$ and each of the
tuples satisfies $cdn()$. They produce $\frake$-trees $e'_1$ and $e'_2$. Because each
tuple has a condition tree $T^{x_1/\gamma_c/\theta_c}$ satisfying
$x_c/\gamma_c/\theta_c=strValu$, by Lemma \ref{lm:updCondi-VS-all-true}, each of $e'_1$
and $e'_2$ has $T^{\calL_c/\theta_c}$ satisfying $\calL_c/\theta_c=strValu$ and each has
a $T^{\calL_t/\theta_t}$. Thus $T^{\calL_t/\theta_t}_1\in e'_1$ and
$T^{\calL_t/\theta_t}_2\in e'_2$ will be updated to $\barT^{\calL_t/\theta_t}_1$ and
$\barT^{\calL_t/\theta_t}_2$ by $\delta V$. $e'_1$ and $e'_2$ become $e_1$ and $e_2$ in
$\delta V(V(S^i))$.

Case (b.1): That $T^{x_1/\gamma_t/\theta_t}_1$ and $T^{x_1/\gamma_t/\theta_t}_2$ are
updated by $\delta S$ and that they appear in different $\frake$-trees mean that there
are two tuples $\tup{x_{1[1]},\ddd}$ and $\tup{x_{1[2]},\ddd}$ where $x_{1[1]}$ and
$x_{1[2]}$ are different bindings of $x_1$, $T^{x_1/\gamma_t/\theta_t}_1\in x_{1[1]}$,
$T^{x_1/\gamma_t/\theta_t}_2\in x_{c[2]}$, and each of tuples satisfies $cdn()$ and
$x_c/\gamma_c/\theta_c=strValu$. $T^{x_1/\gamma_t/\theta_t}_1$ and
$T^{x_1/\gamma_t/\theta_t}_2$ become $\barT^{x_1/\gamma_t/\theta_t}_1$ and
$\barT^{x_1/\gamma_t/\theta_t}_2$ after the update and mapped to
$\barT^{\calL_t/\theta_t}_1$ and $\barT^{\calL_t/\theta_t}_2$ in two different
$\frake$-trees of $V(\delta S(S^i))$. Following the same argument of Case (a),
$\barT^{\calL_t/\theta_t}_1$ and $\barT^{\calL_t/\theta_t}_2$ are in $\delta V(V(S^i))$.

Case (b.2): That $T^{x_1/\gamma_t/\theta_t}_1$ and $T^{x_1/\gamma_t/\theta_t}_2$ are
updated by $\delta S$ and that they appear in a single $\frake$-tree mean that there is
one and only one tuple $\tup{x_1,\ddd}$ where
$T^{x_1/\gamma_t/\theta_t}_1,T^{x_1/\gamma_t/\theta_t}_2 \in x_{1}$. The tuple satisfies
$cdn()$ and there is a tree $T^{x_1/\gamma_c/\theta_c}$ in the tuple satisfying
$x_1/\gamma_c/\theta_c=strValu$. $T^{x_1/\gamma_t/\theta_t}_1$ and
$T^{x_1/\gamma_t/\theta_t}_2$ become $\barT^{x_1/\gamma_t/\theta_t}_1$ and
$\barT^{x_1/\gamma_t/\theta_t}_2$ after the update and mapped to
$\barT^{\calL_t/\theta_t}_1$ and $\barT^{\calL_t/\theta_t}_2$ in a single $\frake$-tree
of $V(\delta S(S^i))$. On the other side, as $T^{x_1/\gamma_t/\theta_t}_1$ and
$T^{x_1/\gamma_t/\theta_t}_2$ are mapped to a single $\frake$-tree $e$ and share the same
condition tree $T^{x_1/\gamma_c/\theta_c}$, $T^{\calL_t/\theta_t}_1$ and
$T^{\calL_t/\theta_t}_2$ share the same condition tree $T^{\calL_c/\theta_c}$ in $e$ and
will be updated by $\delta V$. So $\barT^{\calL_t/\theta_t}_1$ and
$\barT^{\calL_t/\theta_t}_2$ are in the $\frake$-tree of $\delta V(V(S^i))$.

(2) $\delta S$ is minimal

We prove by contrapositive. Let $T^{\calL_t/\theta_t}$ be a tree in the view updated by
$\delta V$. Then from above proofs, $T^{x_1/\gamma_t/\theta_t}$ is updated by $\delta S$
and there exists a tuple $\tup{x_1,\ddd}$ such that $T^{x_1/\gamma_t/\theta_t}$ is in
$x_1$ and $x_1$ has a condition tree $T^{x_1/\gamma_c/\theta_c}$ satisfying ``$cdn()$ and
$x_1/\gamma_c/\theta_c=strValu$".

If $T^{x_1/\gamma_t/\theta_t}$ is not updated by $\delta S'$,  either (a) $x_1$ is not a
variable in the $for$-clause of $\delta S'$, i.e., $x_1$ is not in any tuple and neither
is $T^{x_1/\gamma_t/\theta_t}$, or (b) $x_1$ is in the tuple $\tup{x_1,\ddd}$ but
$T^{x_1/\gamma_t/\theta_t}$ is not in $x_1$, or (c) $x_1$ is in the tuple
$\tup{x_1,\ddd}$ and $T^{x_1/\gamma_t/\theta_t}$ is in $x_1$ but one of ``$cdn()$" and
``$x_c/\gamma_c/\theta_c=strValu$" is not in $\delta S'$.

In Case (a), because $x_1$ is not a variable in $\delta S'$, so
$T^{x_1/\gamma_t/\theta_t}$ will not be updated by $\delta S'$ (this does not prevent
$T^{x_1/\gamma_t/\theta_t}$ from appearing in the view). This means that the
$T^{\calL_t/\theta_t}$ in $V(\delta S'(S^i))$ is different from the
$T^{\calL_t/\theta_t}$ in $\delta V(V(S^i))$ because the assumption assumes that the
$T^{\calL_t/\theta_t}$ in $\delta V(V(S^i))$ is updated. This contradicts the correctness
of $\delta S'$.

In Case (b), because $T^{x_1/\gamma_t/\theta_t}$ is not in $x_1$, so
$T^{x_1/\gamma_t/\theta_t}$ is not in $V(S^i)$. This contradicts the assumption that
$T^{\calL_t/\theta_t}$ is in the view.

In Case (c), if $cdn()$ is violated, the tuple of $T^{x_1/\gamma_t/\theta_t}$ will not be
selected by $V$, so $T^{x_1/\gamma_t/\theta_t}$ is not in $V(S^i)$ which contradicts the
assumption. If $x_1/\gamma_c/\theta_c=strValu$ is violated, $T^{x_1/\gamma_t/\theta_t}$
will not be updated by $\delta V$. This contradicts the assumption that
$T^{\calL_t/\theta_t}$ is updated by $\delta V$.

This concludes that $\delta S$ is a precise translation.
 \\ \done

We note that the theorem gives only a necessary condition but not a sufficient condition.
The reason is that there exists other cases where a view update is translatable. These
will be further presented in the following sections.

\begin{figure}[t] \center
      \includegraphics[scale=0.7]{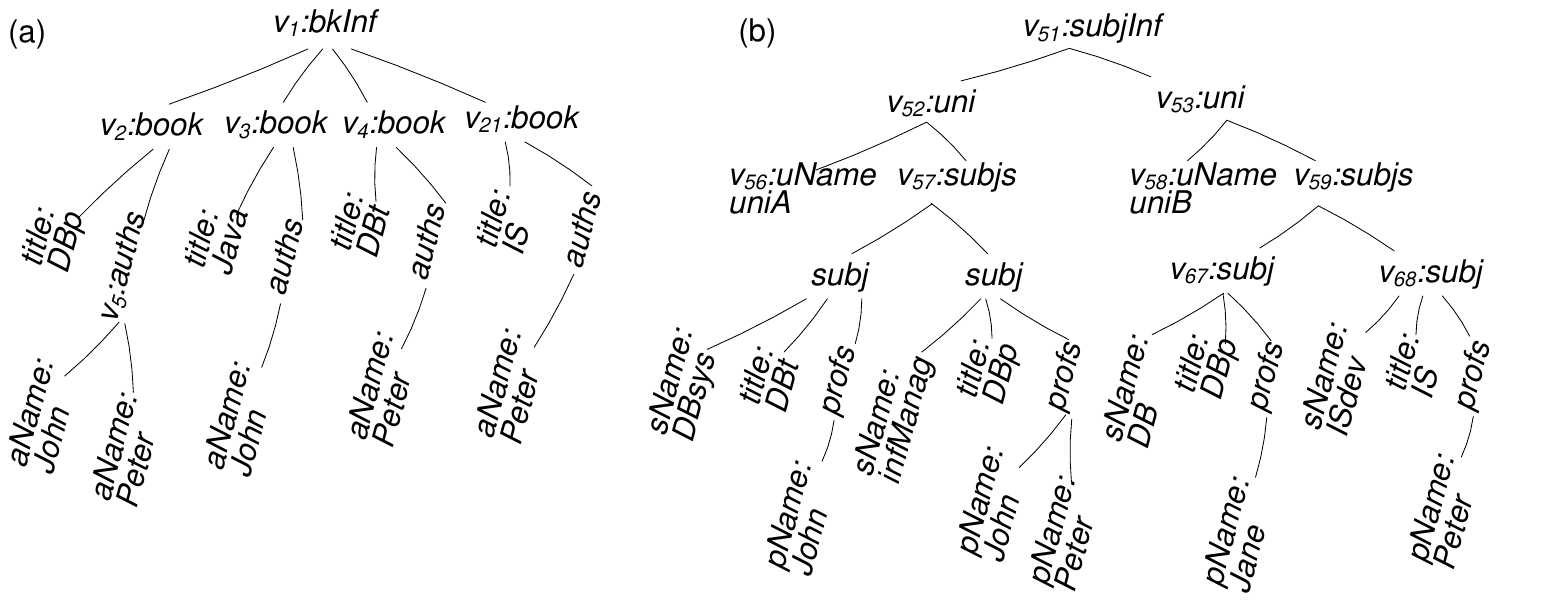}
     \caption{Books and their references}
   \label{f:bkInf}
\end{figure}

We use an example to show how a view update is translated using the results. Figure
\ref{f:bkInf} shows two XML documents. Document (a) stores book information where $auths$
and $aName$ mean authors and author-name elements respectively. Document (b) stores
university subject, textbook and professor information where $uName$, $subjs$, $sName$,
$profs$, and $pName$ mean university-name, subjects, subject-name, professors, and
professor-name respectively.

The view $Qbk$ is defined below to contain, for each use of a book by a university
subject, the author names and the title of the book, the name of the university and the
professors using the book in their teaching.

{\small
\begin{Verbatim}[commandchars=\\\{\},codes={\catcode`$=3\catcode`^=7\catcode`_=8}]
 <Qbk>\{  for x in doc("bkInf.xml")/bkInf/book,
             y in doc("subjInf.xml")/subjInf/uni,
             z in y/subjs/subj
         where x/title=z/title
         return <use>\{x/auths\}\{x/title\}\{y/uName\}\{z/profs\}</use>
 \}</Qbk>
\end{Verbatim}
}

\begin{figure}[h] \center
      \includegraphics[scale=0.7]{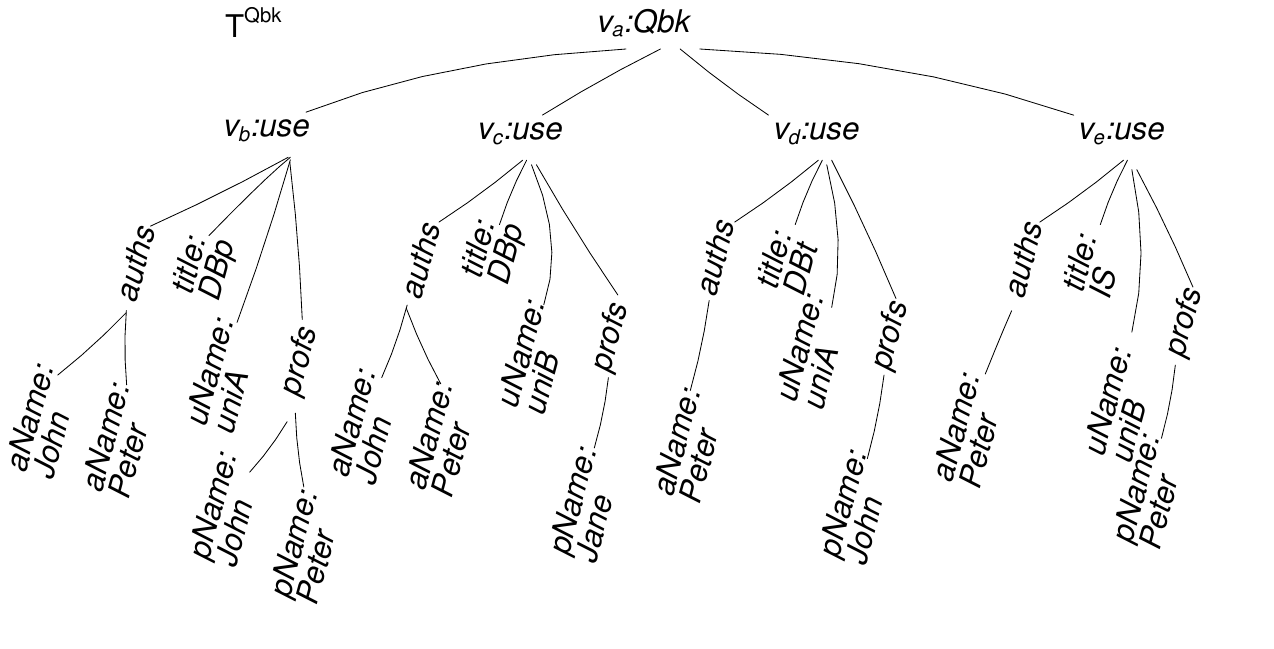}
     \caption{author-books and universities using them}
   \label{f:bkUse}
\end{figure}

The view instance for the XML documents is shown in Figure \ref{f:bkUse}.

Now assume that the user of the view wants to add author $Susan$ to the textbook $IS$ in
the view using the update statement below.

{\small
\begin{Verbatim}[commandchars=\\\{\},codes={\catcode`$=3\catcode`^=7\catcode`_=8}]
 for r in view(Qbk)/Qbk/use
 where r/title="IS"
 update r/auths \{ insert <aName>Susan</aName>\}
\end{Verbatim}
}

\begin{figure}[h] \center
      \includegraphics[scale=0.7]{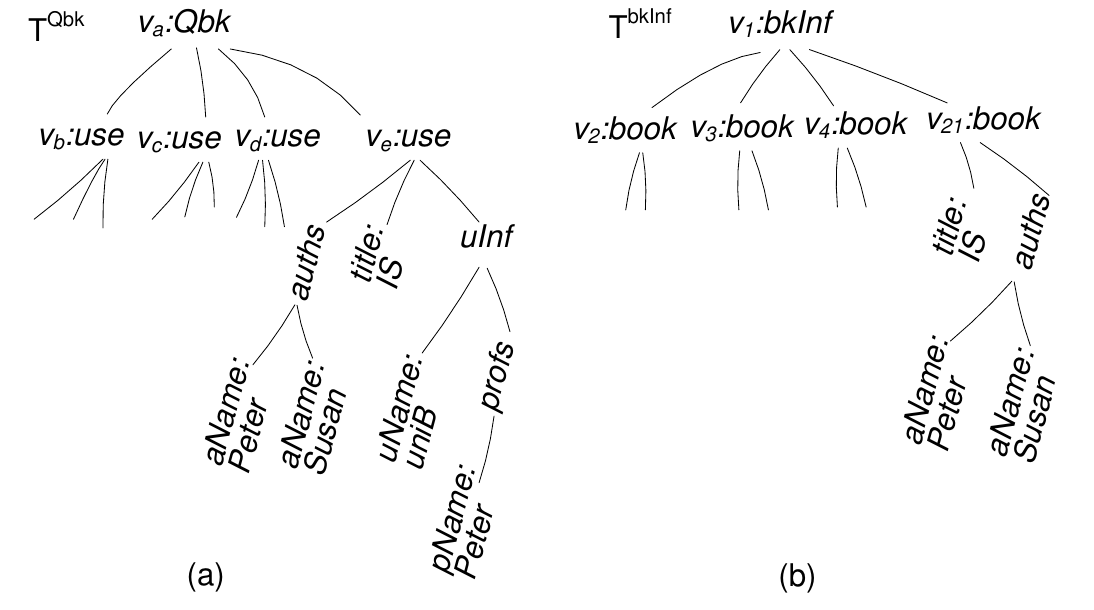}
     \caption{An insertion update}
   \label{f:insAuth}
\end{figure}

With this statement, the user expects that next time when the view is selected, the
output is Figure \ref{f:insAuth}(a) where trees $v_b$, $v_c$ and $v_d$ are the same as
those of Figure \ref{f:bkUse} and tree $v_e$ contains the newly added author $Susan$.

In the update statement, the update condition path and the update target path are
$r/title$ and $r/auths$. The full view paths of the two paths are: $Qbk/use/title$ and
$Qbk/use/auths$. In the paths, $Qbk$ is $v$ of Formula (\ref{eq:vpath}), $use$ is
$\mathfrak{e}$, $title$ is $\calL_c$, $auths$ is $\calL_t$, and $\theta_c$ and $\theta_t$
are $\phi$. Following Procedure \ref{def:mapping} by using $title$ and $auths$, we find
the expressions $x/title$ and $x/auths$. By Algorithm \ref{alg:translate}, the following
source update is derived:

{\small
\begin{Verbatim}[commandchars=\\\{\},codes={\catcode`$=3\catcode`^=7\catcode`_=8}]
 for x in doc("bkInf.xml")/bkInf/book,
     y in doc("subjInf.xml")/subjInf/uni,
     z in \$y/subjs/subj
 where x/title=z/title and x/title="IS"
 update x/auths \{ insert <aName>Susan</aName>\}
 \end{Verbatim}
}

When this statement is executed against Figure \ref{f:bkInf}(a), the document becomes
Figure \ref{f:insAuth}(b) where trees $v_2$, $v_3$ and $v_4$ are the same as those in
Figure \ref{f:bkInf}(a) and $v_{21}$ is changed. The view instance will appear as
expected by the user when selected next time.


 \vspace{-1ex}

\section{Update Translation when ${\cal L}_t/\theta_t \not=\phi$ and $x_c\not=x_t$} \label{sec:translation2}
 \vspace{-1ex}
We look into the translation problem when the mappings of the update condition path and
the update target path are led by different variables. The results of this section
generalize the view update problem in the relational views when they are defined with the
join operator.

In general, \emph{view updates are not translatable in the case of $x_c\not=x_t$}.
Consider two tuples where the binding $x_t$ is copied to $x^{(1)}_t$ and $x^{(2)}_t$ to
combine with two bindings  $x_{c[1]}$ and $x_{c[2]}$ of $x_c$ by the context-based
production as
   \[\tup{\ddd,x^{(1)}_t,\ddd,x_{c[1]},\ddd}\]
   \vspace{-1.5em}
   \[\tup{\ddd,x^{(2)}_t,\ddd,x_{c[2]},\ddd}\]
Assume that in the view, the update condition $x_c/\gamma_c/\theta_c$ is satisfied in
$x_{c[1]}$ by violated in $x_{c[2]}$.  Then, the copy of $x_t$ corresponding to the first
tuple will be updated but the one to the second tuple will not. In the source, if $x_t$
is updated, not only the first copy of $x_t$ changes, but also the second copy. In other
words, the translated source update has view side-effect. However, if $x_t$ in the source
is not updated, all its copies in the view will not be changed.

Although generally view updates, when $x_c\not=x_t$, are not translatable, for the
following view update, a precise translation exists. \\

 \begin{equation}\label{eq:join-V}\end{equation}
 \vspace*{-4em}
\begin{Verbatim}[commandchars=\\\{\},codes={\catcode`$=3\catcode`^=7\catcode`_=8}]
 $V$:   \tago{$v$}\{ for $x_1$ in $p_1$, \ \ $\ddd$, \ \ $x_n$ in $p_n$
             where  $\ddd$ and $x_c/\gamma_c/\theta_c=x_{c+1}/\gamma_{c+1}/\theta_{c+1}$ and $\ddd$
             return $rtn(x_1,\ddd,x_n)$  \}\tago{$/v$}
\end{Verbatim}
where $x_c/\gamma_c$ is in $rtn(x_1,\ddd,x_n)$, i.e., $x_c/\gamma_c/\theta_c$ is exposed in the view. \\

 \noindent $\delta V$:
 \begin{equation}\label{eq:join-dV}(\bar{p_s}, \ \ v/\mathfrak{e}/{\cal L}_c/\theta_c=strValu,
  \ \ v/\mathfrak{e}/{\cal L}_t/\theta_t,\ \
 del(T)|ins(T))\end{equation}
where $x_t$ is either  $x_{c}$ or $x_{c+1}$. \\

The condition requires that, in the view definition, $x_c/\gamma_c$ must be a front part
of one of the join path $x_c/\gamma_c/\theta_c$. At the same time, the path in view
mapped from $x_c/\gamma_c/\theta_c$ must be the update condition path. Furthermore, the
mapping of the update target path must be led by the same variable $x_c$ leading the
update condition path or by the variable $x_{c+1}$ that joins $x_c$ in the view
definition.

Consider Example \ref{ex:view-def}. With the condition, $y/D=z$ and $z=``1"$, in the
$where$ clause, for a view update to be translatable, the mapping $x_c/\gamma_c/\theta_c$
of the view update condition path must be $z$ or $y/D$, and the mapping
$x_t/\gamma_t/\theta_t$ of the view update target path must be ended with $F$, $G$ or
$E$. We note that if $x_t/\gamma_t/\theta_t$ is ended with $C$ or $H$, then
$x_t/\gamma_t/\theta_t$ is a prefix of one of the paths in the join condition and the
update will not be translatable.

\begin{theo}\label{th:join-precise}
Given the view $V$ and a view update $\delta V$ defined above, update $\delta S$ of
Formula (\ref{eq:deltaS}) is a precise translation of the view update $\delta V$ if (i)
${\cal L}_t/\theta_t \not=\phi$, and (ii) $x_c/\gamma_t/\theta_t$ does not proceed any
path in the $where$ clause of $\delta S$.
\end{theo}

\renewcommand{\Ttr}{\ensuremath{T^{x_t/\gamma_t}}}  
\renewcommand{\Tcr}{\ensuremath{T^{x_c/\gamma_c}}}  
\renewcommand{\Tts}{\ensuremath{T^{x_t/\gamma_t/\theta_t}}}  
\renewcommand{\Ttv}{\ensuremath{T^{\calL_t/\theta_t}}}  
\renewcommand{\Tcs}{\ensuremath{T^{x_c/\gamma_c/\theta_c}}}  
\renewcommand{\Tcv}{\ensuremath{T^{\calL_c/\theta_c}}}  
\newcommand{\bTtv}{\ensuremath{\bar{T}^{\calL_t/\theta_t}}}  
\newcommand{\bTts}{\ensuremath{\bar{T}^{x_t/\gamma_t/\theta_t}}}  

\noindent \bfit{Proof.} The notation of this proof follows that of the proof for Theorem
\ref{th:typic-precise} and Figure \ref{f:tup-e}. Consider two tuples
$t_1=\tup{x_t^{(1)},x_{c[1]},\ddd}$ and $t_2=\tup{x_t^{(2)},x_{c[2]},\ddd}$ in the
evaluation of $\delta S(S)$ where $x_t^{(1)}$ and $x_t^{(2)}$ are copies of $x_t$ and
$x_{c[1]}$ and $x_{c[2]}$ can be the same. If one is updated by $\delta S$, the other is
updated too. The reason is that for $\Tts_1\in x_t^{(1)}$ and $\Tts_2\in x_t^{(2)}$,
because of the join condition in Formula \ref{eq:join-V}
$x_c/\gamma_c/\theta_c=x_{c+1}/\gamma_{c+1}/\theta_{c+1}$ and because $x_{c+1}=x_t$ and
$x_t^{(1)}=x_t^{(2)}$, a condition tree $\Tcs_1$ exists for $\Tts_1$ and $\Tcs_2$ exists
for $\Tts_2$ and $\Tcs_1= \Tcs_2$. Consequently if $\Tcs_1$ satisfies the update
condition, so does $\Tcs_2$. So either both $\Tts_1$ and $\Tts_2$ are updated or none is
updated. Following Lemma \ref{lm:VS-chged2same}, if $e_1$ and $e_2$ are mapped from
$\Tts_1$ and $\Tts_2$ respectively, if one is updated, the other is updated too.

The remaining proof can be completed by following the argument of the proof of Theorem
\ref{th:typic-precise}. \done

\section{Update Translation when ${\cal L}_t/\theta_t =\phi$} \label{sec:translation3}
 \vspace{-1ex}

In this section, we identify translatable cases where ${\cal L}_t/\theta_t =\phi$, that
is, the update target path is $v$ or $v/\frake$. In the case of $v$, the update itself is
an addition or a removal of an $\frake$-tree. In the case of $v/\frake$, the update is an
insertion or a deletion of a $\gamma$-tree.

\begin{figure*}[t] \center
      \includegraphics[scale=0.8]{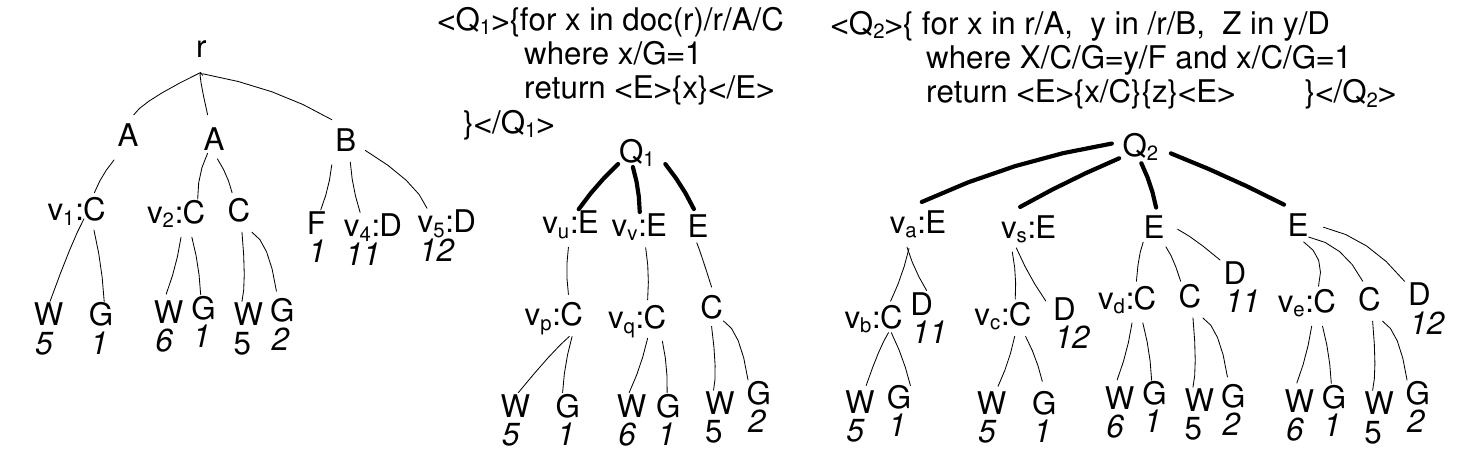}
     \caption{Two views to show updates to $E$ and to $Q$}
   \label{f:basic}
\end{figure*}

Obviously if the user does not know the structure of the view, wrong subtrees can be
added. As an example, consider $Q_1$ in Figure \ref{f:basic}. The path $Q_1/E$ allows
child elements labeled with $C$. If the user adds a sub tree labeled with $F$ under
$v_u$, the update violates the view definition. We exclude this type of cases and assume
that the user knows the structure of the view and the updates aim to maintain such a
structure.

In general, insertion updates are not translatable when ${\cal L}_t/\theta_t =\phi$. A
number of reasons exist for this. The first is that there is no unique way to apply
insertions to the source documents in many cases. The second reason is that the updates
violate the context-based production. The third reason is there is no way for the user to
write an update statement with a specific enough condition to update the view while the
context-based production is not violated. We use three examples to illustrate the
reasons.

\begin{exam}
Consider $Q_1$ in Figure \ref{f:basic}. If another subtree $(E\ (C \ (W:2)(G:8)))$ is
inserted to $Q_1$, in the source the subtree $(C \ (W:2)(G:8))$ needs to be inserted to
$r$. We cannot find a unique way to do so as the subtree can be inserted to an existing
$A$ element or a new $A$ element is created and the subtree is inserted under the new $A$
element. \end{exam}

\begin{exam}
 Consider $Q_1$ in Figure \ref{f:basic} again. If an update is an insertion of $(C \
(W:2)(G:8))$ under $v_u$, the context-based production is violated. By the context-based
production, if $x$ in the $return$ clause is not followed by any path expression, only
one $C$ element is allowed in each $E$ tree.
\end{exam}

\begin{exam}
 Consider $Q_2$ in Figure \ref{f:basic} where $C$ elements are selected by
$x/C$ in the $return$ clause. If the user wants to insert another $C$ element under both
$v_a$ and $v_s$ (but not the other $E$ elements) such that the context-based production
is satisfied, the user has no way to specify an accurate condition for this because the
node identifiers, $v_a$ and $v_s$, are not available to the user.
\end{exam}

For the same reasons, many deletion updates are not translatable. However in the case
where all the expressions in the $return$ clause start with the same variable, deletion
updates to such views are translatable. We show the details below.

Let the view definition be
 \vspace*{1em}
  \begin{equation}\label{eq:viewE}\end{equation}
 \vspace*{-4.5em}
\begin{Verbatim}[commandchars=\\\{\},codes={\catcode`$=3\catcode`^=7\catcode`_=8}]
  $V$: \tago{$v$}\{ for $x_1$ in $p_1$, $\ddd$, $x_n$ in $p_n$
         where  $cdn(x_1,\ddd,x_n)$
         return $rtn(x_1)$  \}\tago{$/v$}
\end{Verbatim}

In the view, only the variable $x_1$ is involved in the $return$ clause. Let the update
statement to the view be
 \vspace*{1em}
  \begin{equation}\label{eq:viewE-dV}\end{equation}
 \vspace*{-4.5em}
\begin{Verbatim}[commandchars=\\\{\},codes={\catcode`$=3\catcode`^=7\catcode`_=8}]
  $\delta$$V$: for $e$ in $v/\frake$
     where $e/\calL_c/\theta_c=aVal$
     update $e$ (delete $\calL_t$)
\end{Verbatim}

The translated source update is
 \vspace*{1em}
  \begin{equation}\label{eq:viewE-dS}\end{equation}
 \vspace*{-4.5em}
\begin{Verbatim}[commandchars=\\\{\},codes={\catcode`$=3\catcode`^=7\catcode`_=8}]
  $\delta$$S$: for $x_1$ in $p_1$, $\ddd$, $x_n$ in $p_n$
     where  $cdn(x_1,\ddd,x_n)$ and $x_1/\gamma_c/\theta_c=aVal$
     update $x_1/\gamma_t/..$ (delete $\calL_t$)
\end{Verbatim}
In the formulae, $\calL_t$ is the last element of $x_1/\gamma_t$. To allow a $\calL_t$
node to be inserted to or deleted from the source document, the target path must be
$x_1/\gamma_t/..$ .

\begin{theo}
Given the view definition $V$, the source update $\delta S$ is a precise translation of
the view update $\delta V$ if $x_1/\gamma_t/..$ does not proceed any of the paths in the
$where$ clause of $\delta S$.
\end{theo}

\renewcommand{\Tts}{\ensuremath{T^{x_t/\gamma_t/\theta_t}}}  
\renewcommand{\Ttv}{\ensuremath{T^{\calL_t/\theta_t}}}  
\renewcommand{\Tcs}{\ensuremath{T^{x_c/\gamma_c/\theta_c}}}  
\renewcommand{\Tcv}{\ensuremath{T^{\calL_c/\theta_c}}}  

{\bf Proof}: We follow Definition \ref{def:precise} to prove $V(\delta S(S^i))=\delta
V(V(S^i))$ and omit the proof that $\delta S$ is minimal. We note that
$\calL_t\not=\calL_c$ implies $x_1/\gamma_c\not=x_1/\gamma_t$.

 $\subseteq:$ \ Assume that $e'_1$ and $e'_2$ are two $\frake$-trees in $V(\delta S(S^i))$.
Then there exists two tuples $t'_1=<\bar{x}_1^{(1)},\ddd>$ and
$t'_2=<\bar{x}_1^{(2)},\ddd>$ for $e'_1$ and $e'_2$ and they satisfy $cdn()$ of $V$. That
the two tuples are updated by $\delta S$ means that they are the results of updating two
tuples $t_1=<x_1^{(1)},\ddd>$ and $t_2=<x_1^{(2)},\ddd>$ by $\delta S()$ and $t_1$ and
$t_2$ satisfy $cdn()$ and have condition trees $T^{x_1/\gamma_c/\theta_c}_1$ and
$T^{x_1/\gamma_c/\theta_c}_2$ satisfying $x_1/\gamma_c/\theta_c=aVal$, and the update
deletes trees like $T^{x_1/\gamma_t}$. Consequently $T^{\calL_t}$s are not in $e'_1$ and
$e'_2$.

On the other side, as $t_1$ and $t_2$ satisfies $cdn()$, they produce $e_1$ and $e_2$ in
$V(S^i)$. At the same time, $e_1$ and $e_2$ have condition trees $T^{\calL_c/\theta_c}_1$
and $T^{\calL_c/\theta_c}_2$ satisfying $\calL_c/\theta_c=aVal$ (Lemma
\ref{lm:updCondi-VS-all-true}), they are updated as $T^{\calL_t}$s will be deleted from
from them. So they become $e'_1$ and $e'_2$ and are in $\delta V(V(S^i))$.

 $\supseteq:$ \ Let $e'_1$ and $e'_2$ be $\frake$-trees in $\delta V(V(S^i))$.
Then there exist $e_1$ and $e_2$ in $V(S^i)$ and $\delta V$ deletes $T^{\calL_t}$s from
them. That is, $e_1$ and $e_2$ have condition trees satisfying $cdn()$ and
$\calL_c/\theta_c=aVal$. $e_1$ and $e_2$ are for two tuples $t_1=<x_1^{(1)},\ddd>$ and
$t_2=<x_1^{(2)},\ddd>$ in $V$ and the two tuples satisfy $cdn()$.

On the other side, $t_1$ and $t_2$ satisfy $cdn()$ and $x_1/\gamma_c/\theta_c=aVal$
(Lemma \ref{lm:updCondi-VS-all-true}), they will be updated and $T^{\calL_t}$s will be
deleted from them. So because of Lemma \ref{lm:noChange2cdn}, they become
$t'_1=<\bar{x}_1^{(1)},\ddd>$ and $t'_2=<\bar{x}_1^{(2)},\ddd>$. When $\delta S(S^i)$ is
evaluated against $V$, $t'_1$ and $t'_2$ produces $e'_1$ and $e'_2$ which do not contain
any $T^{\calL_t}$s. So they are in $V(\delta S(S^i))$

 $\delta S$ is minimal: \ If a tree is not relevant to the view, the
tree does not satisfy $cdn(x_1,\ddd,x_n)$ and it will not be updated by $\delta S$. \done
\\

For the same view definition $V$ in (5), if the update is applied to the root node as the
following,
 \vspace*{1em}
  \begin{equation}\label{eq:viewRoot-dV}\end{equation}
 \vspace*{-4.5em}
\begin{Verbatim}[commandchars=\\\{\},codes={\catcode`$=3\catcode`^=7\catcode`_=8}]
$\delta$$V$:  for $u$ in $v$,
     where $u/\frake/\calL_c/\theta_c=aVal$
     update $u$ (delete $\frake$)
\end{Verbatim}
 the translated soruce update is
 \vspace*{1em}
  \begin{equation}\label{eq:viewRoot-dS}\end{equation}
 \vspace*{-4.5em}
\begin{Verbatim}[commandchars=\\\{\},codes={\catcode`$=3\catcode`^=7\catcode`_=8}]
$\delta$$S$:  for $x_1$ in $p_1$, $\ddd$, $x_n$ in $p_n$
     where  $cdn(x_1,\ddd,x_n)$ and $x_1/\gamma_c/\theta_c=aVal$
     update $x_1/..$ (delete $L(x_1)$)
\end{Verbatim}

We note that when an $\frake$ node is deleted, deleting all the $\gamma$ trees from their
parent nodes in the source document is not enough. The binding of the variable must be
deleted.

\begin{theo}
Given view definition $V$ in Formula (\ref{eq:viewE}), the source update $\delta S$ in
Formula (\ref{eq:viewRoot-dS}) is a precise translation of the view update $\delta V$ in
Formula (\ref{eq:viewRoot-dV}).
\end{theo}

\emph{proof}: \ Let $t_1=<x_1^{(1)},\ddd>$ and $t_2=<x_1^{(2)},\ddd>$ be two tuples in
$fortup(V)$, $x_1^{(1)}$ and $x_1^{(2)}$ be two copies of $x_1$ in the source, $e_1$ and
$e_2$ be two $\frake$-trees for the tuples in $V(S^i)$, and $e_1$ and $e_2$ are deleted
by $\delta V$. Because $e_1$ and $e_2$ are in $V(S^i)$, $t_1$ and $t_2$ satisfy $cdn()$.
$e_1$ and $e_2$ being deleted by $\delta V$ means that each of them has a subtree
$T^{\calL_c/\theta_c}$ satisfying $\calL_c/\theta_c=aVal$. By Lemma
\ref{lm:updCondi-VS-all-true}, each of $t_1$ and $t_2$ has a tree
$T^{x_1/\gamma_c/\theta_c}$ satisfying $x_1/\gamma_c/\theta_c=aVal$. Thus $t_1$ and $t_2$
will be updated by $\delta S$ meaning the binding of $x_1$ will be deleted from the
source. Consequently $t_1$ and $t_2$ will not be in $fortup(V(\delta S())$ and $e_1$ and
$e_2$ will not be in $V(\delta S())$.

The proof that $\delta S$ is minimal is similar to that of Theorem
\ref{th:typic-precise}.  \done

 \vspace{-1em}
\section{Conclusion}\label{sec:conclusion}
 \vspace{-1em}
 In this paper, we defined the view update problem in XML and shown the
factors determining the translation problem. We identified the cases where view updates
are translatable, shown a translation algorithm, gave the translated source updates, and
proved the source updates are precise.

The translatability of view updates is information dependent. In this paper, we assume
the only information available is the view definition and the update. When other
information like keys and references are used in the translation, different algorithms
and different source updates may be obtained. We leave the investigation of these
problems as future work.

 \vspace{-1em}
\bibliography{../../../hd_refs/refs}
\bibliographystyle{plain}

\end{document}

%% file: jl_definitions.tex
\usepackage{graphicx} 
\usepackage{latexsym}
\usepackage{amssymb}

\usepackage{amsthm}
\usepackage{verbatim}
\usepackage{fancyvrb}

\newcommand{\done}{\hspace{1em}$\Box$}

\newcommand{\bfit}[1]{\textbf{\textit{#1}}\index{#1}} 

\newcommand{\ddd}{\cdots}

\renewcommand{\em}[1]{\textbf{#1}}
\renewcommand{\emph}[1]{\textbf{#1}}
\renewcommand{\bfit}[1]{\textbf{\textit{#1}}}

\newlength{\theoskip}
\newlength{\theoskipAlgo}
\newtheorem{lemm1}{Lemma}
\newtheorem{theo1}{Theorem}
\newtheorem{coro1}{Corollary}
\newtheorem{proposi1}{Proposition}
\newtheorem{obsv1}{Observation}
\theoremstyle{remark}  
\newtheorem{remk1}{Remark}[section]
\theoremstyle{definition} 
\newtheorem{exam1}{Example}
\newtheorem{defi1}{Definition}
\newtheorem{algo1}{Algorithm}
\newtheorem{func1}{Function}
\newtheorem{query1}{Query}
\newtheorem{expl1}{Pf\hspace*{-.8ex}}
\newtheorem{proc1}{Procedure}

\newenvironment{lemm}{\vspace{\theoskip}\begin{lemm1}}{\end{lemm1}}

\newenvironment{theo}{\vspace{\theoskip}\begin{theo1}}{\end{theo1}}

\newenvironment{defi}{\vspace{\theoskip}\begin{defi1}}{\end{defi1}}
\newenvironment{exam}{\vspace{\theoskip}\begin{exam1}}{\end{exam1}}

\newenvironment{proc}{\vspace{\theoskip}\begin{proc1}}{\end{proc1}}

\newcounter{mylisti}
\newcounter{mylista}
\newcounter{mylist1}

\newenvironment{list1}{\begin{list}{(\arabic{mylist1})}
   {\usecounter{mylist1} \itemsep=0ex \topsep=0mm
             \parsep=0ex \parskip=0ex \partopsep=0mm}}
   {\end{list}}

\newenvironment{itm}{\begin{list}{$\centerdot$}
  { \itemsep=0ex \topsep=0mm \parsep=0ex \parskip=0ex \partopsep=0mm}}
   {\end{list}}